\documentclass[12pt]{article}
\usepackage{epsfig,rotate,wasysym}
\usepackage{natbib}
\usepackage{xcolor}
\usepackage{ulem}
\usepackage{framed}

\usepackage{titlesec}
\titlespacing*{\section}{0.0\baselineskip}{0.5\baselineskip}{0.5\baselineskip}

\usepackage[disable]{todonotes}
\usepackage{pdfpages}
\title{Detecting the elemental and molecular signatures of life:
Laser-based mass spectrometry technologies}

\usepackage{times}  

\begin{document}

\newcommand\bra[1]{{<#1|}}
\newcommand\ket[1]{{|#1>}}
\def\window{Si$_3$N$_4$} 
\newif\ifincludefigures
\newif\ifnofigures

\def\etal{{\em {\em {\em et al.}}}}

\includefigurestrue

\pagestyle{myheadings}
\markright{Laser-based mass spectrometry technologies}

\thispagestyle{empty}

\addtocounter{page}{-1}
\pagenumbering{roman}
\pagenumbering{arabic}

\addtocounter{page}{-1}

\clearpage

\noindent {\Large \bf Detecting the elemental and molecular signatures of life: \\ Laser-based mass spectrometry technologies} \ \\ \ \\

\noindent {\large Niels F.W. Ligterink\textsuperscript{1}, Andreas Riedo\textsuperscript{2}, Marek Tulej\textsuperscript{2}, Rustam Lukmanov\textsuperscript{2}, Valentine Grimaudo\textsuperscript{2}, Coenraad de Koning\textsuperscript{2}, Peter Wurz\textsuperscript{2}, Christelle Briois\textsuperscript{3}, Nathalie Carrasco\textsuperscript{4}, Ricardo Arevalo Jr.\textsuperscript{5}, and William B. Brinckerhoff\textsuperscript{6} \\
\\
\textsuperscript{1}Center for Space and Habitability, University of Bern, Switzerland \\
(niels.ligterink@csh.unibe.ch) \\
\textsuperscript{2}Space Research and Planetary Sciences, Physics Institute, University of Bern, Switzerland \\
\textsuperscript{3}Laboratoire de Physique et Chimie de l’Environnement et de l’Espace (LPC2E), Orléans, France \\
\textsuperscript{4}Laboratoire Atmosphères, Milieux et Observations Spatiales (LATMOS), Guyancourt, France \\ 
\textsuperscript{5}Department of Geology, University of Maryland, USA \\
\textsuperscript{6}NASA Goddard Space Flight Center, USA
\\
\\
\noindent A whitepaper written for the Planetary Science and Astrobiology Decadal Survey 2023--2032 of the National Academy of Sciences.

\section*{Abstract} 

The identification of extraterrestrial life is one the most exciting and challenging endeavors in space research. The existence of extinct or extant life can be inferred from biogenic elements, isotopes, and molecules, but accurate and sensitive instruments are needed to detect these species. In this whitepaper we show that Laser-based Mass Spectrometers are promising instrument for the in situ identification of atomic, isotopic, and molecular biosignatures. An overview of Laser ablation/Ionization Mass Spectrometry (LIMS) and Laser Desorption/Ionization Mass Spectrometry (LD-MS) instruments developed for space exploration is given. Their uses are discussed in the context of a Mars scenario and an Europa scenario. We show that Laser-based Mass Spectrometers are versatile and technologically mature instruments with many beneficial characteristics for the detection of life. Future planetary lander and rover missions should be encouraged to make use of Laser-based Mass Spectrometry instruments in their scientific payload.

\clearpage

\textbf{Recommendation:} This whitepaper highlights the advantages that laser-based mass spectrometry offers for the in situ detection of chemical biosignatures on planets and moons. Future planetary lander and rover missions should be encouraged to make use of laser-based mass spectrometry instruments in their scientific payload. 


\section{Introduction}

The detection and identification of signatures of life, extinct or extant, on Solar System bodies other than Earth, is nowadays one of the most challenging endeavors in space research. These signatures can be abundances of specific biogenic elements, fractionated isotopes due to metabolic processes, or biomolecules, such as amino acids, nucleobases, and more complex macromolecular networks \citep{hays2017}. Past in situ space exploration missions have attempted to identify biosignatures, albeit with limited success. Some problems have been pointed out with traditional planetary exploration instruments, such as insufficient sensitivity, the possibility that faint biosignatures in micrometer-sized fossils get lost in bulk analysis techniques, or the chemical alteration of biosignatures during, e.g., sample processing. To counter these issues, a new type of analysis technique is required, which combines high detection sensitivity and spatial resolution measurement capabilities at micrometer level and below, and does not chemically alter the sample material of interest. An analysis technique that combines all these requirements is laser-based mass spectrometry, including Laser ablation Ionization and Laser Desorption Mass Spectrometry (LIMS, LD-MS). LIMS and LD-MS instruments have the advantage that they do not require carrier gasses, energy-consuming ovens, or extensive sample preparation steps. This sets them apart from to date traditional planetary exploration instruments, such as pyrolysis Gas Chromatography – Mass Spectrometry (pyr GC – MS) or other chromatographic methods involving extraction and preparation of the sample by wet or gaseous chemistry, biasing the analysis to certain chemical groups, before mass spectrometric analysis. It also means that LIMS and LD-MS instruments can be significantly more compact, lower in weight and power consumption, and are less complex. In this whitepaper, we give an overview of prototype LIMS and LD-MS systems for space exploration, their versatility, and their beneficial characteristics for future deployment for the detection of life. A Mars and a Europa scenario are used to show how the instruments that are developed for use on planetary science missions, can operate on rocky planets or ice-covered ocean worlds, respectively, and search for signatures of life. Finally, we will discuss other uses of LIMS and LD-MS instruments on future space exploration missions.
 

\section{Laser-based mass spectrometry measurement techniques, LIMS and LD-MS}

In its essence, all LIMS and LD-MS instruments operate following the same principle. A single laser pulse is fired at a solid material, where the laser-matter interaction results in the release of atoms or molecules to the gas-phase. During the same laser-matter interaction a fraction of the species is ionized, which makes it possible to detect and identify them using a mass analyzer. The main difference between LIMS and LD-MS is their applied laser irradiance (W cm$^{-2}$), which induces material ablation or desorption and ionization, respectively. By tuning the laser irradiance, which in turn depends on parameters such as laser pulse length, laser wavelength, focal spot size, and pulse energy, different laser-matter interactions can be achieved, which can be grouped in both laser ablation and laser desorption conditions outlined in the following in more detail. \\

\textbf{Laser Desorption Mass Spectrometry (LD-MS).} At moderate laser power densities (MW to GW/cm$^{2}$) this interaction results in laser desorption; the breaking of physisorbed bonds of molecules to a surface, followed by the intact release of these molecules to the gas phase. Ionization is either realized directly at the desorption of molecules from the surface or via a molecule-laser interaction in the gas-phase. Typically, nanosecond laser systems are preferred for LD-MS, as, in comparison to short-pulsed femtosecond laser systems, the photon density is significantly lower, which induces only modest fragmentation of the molecules desorbed from the surface; a severe fragmentation would challenge their identification. Laser desorption has a number of advantages: in contrast to pyrolysis, it does not chemically alter molecules or induce reactions between organic molecules and salts or minerals, although compound molecules may form. At mild desorption conditions near the laser desorption threshold, molecular fragmentation is limited, which prevents isobaric interference and contrasts with, e.g., electron ionization, which results in extensive fragmentation. \\

\textbf{Laser Ablation Ionization Mass Spectrometry (LIMS).} At high laser irradiances (GW/cm$^{2}$ to TW/cm$^{2}$ level), the laser-matter interaction results in laser ablation conditions; a volatile matter-plasma conversion, accompanied by the near-complete atomization and, to a large extent, ionization of the material. Laser ablation is particularly well suited to investigate the element and isotope composition of solids of interest. By applying multiple laser pulses to the same surface spot, layers of material can be removed shot by shot, and subsurface material can be analyzed (chemical depth profiling). In particular femtosecond laser pulses at UV wavelengths are well suited for this because of efficient ionization, close to stoichiometric ion production, and minimized element fractionation effects. With well-focused laser beams, focal spot sizes of several micrometers can be achieved, which makes it possible to perform chemical analysis of micrometer-sized features (microfossils, finely laminated biopatterns, etc). The application of chemical imaging and depth profiles in close vicinity to each other allows, e.g., spatially resolved chemical analysis of the solids in three dimensions, and, e.g., to identify regions of interest. The laser ablation process can also help by the identification of intact neutral molecules when neutral molecules form clusters with metal ions in the plasma. \\

Subsequent analysis of the ions formed in the laser desorption or ablation process is performed with a mass analyzer. For in situ space applications, several miniature mass analyzers have been developed, such as Quadrupole Mass Spectrometers (QMS) \citep{li2017,goesmann2017} high mass resolution Orbitrap systems \citep{briois2016,arevalo2019,willhite2020}, and time-of-flight Mass Spectrometers (TOF MS) \citep{rohner2003,rohner2004}, some of which are optimized for laser ablation and laser desorption studies. Although in principle not necessary for LIMS and LD-MS systems, the mass analyzer can be outfitted with a post ionization stage, such as electron ionization or a secondary laser ionization technique \citep{getty2012,anderson2012} to ionize neutral species and significantly increase the systems’ sensitivity, however, at the expense of higher complexity of the overall system (e.g., more optical elements).


\begin{figure}[h!]
\centering 
\includegraphics[width=1.0\columnwidth]{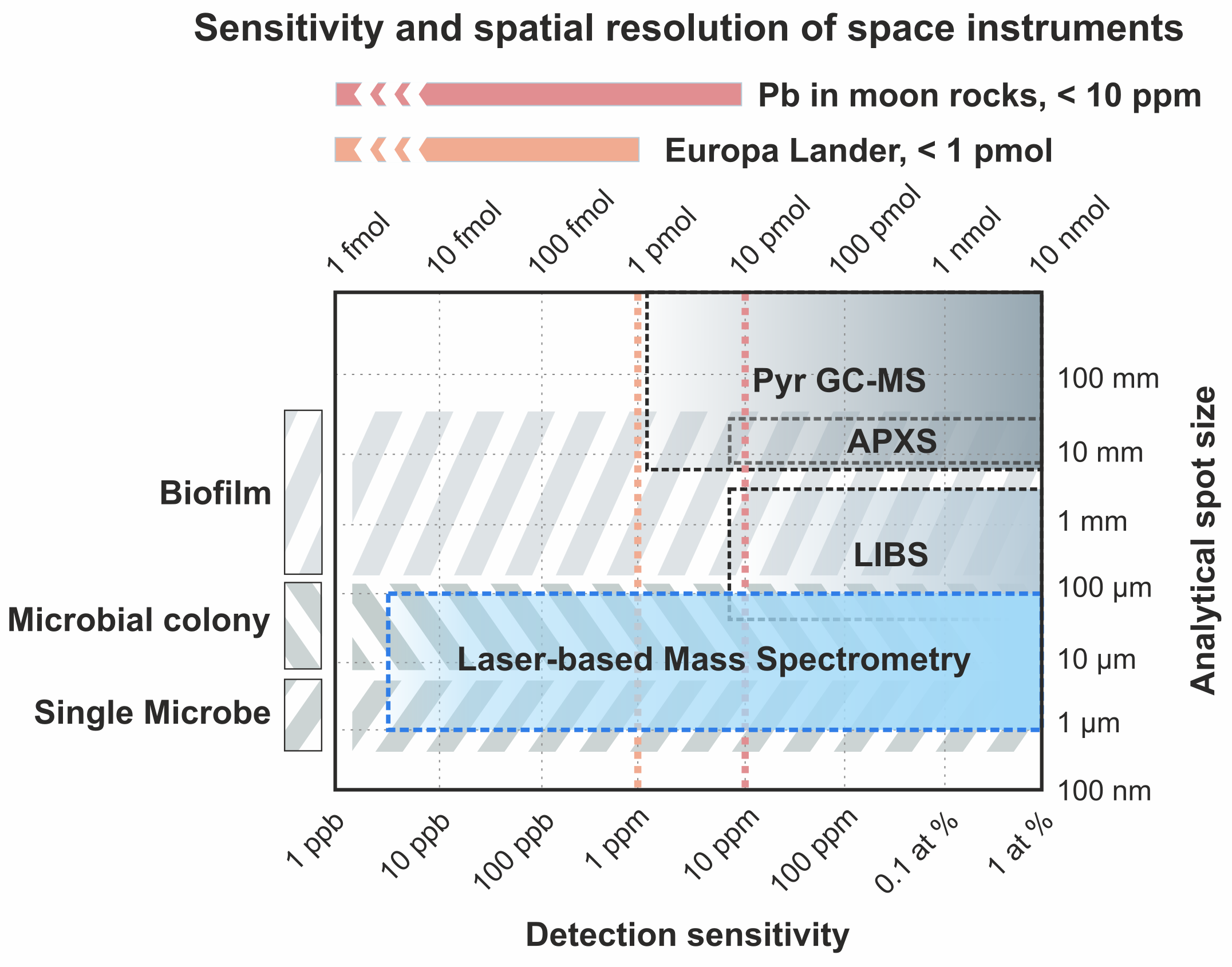} 
\caption[]{Sensitivity and spatial resolution of elemental and chemical analysis instruments developed for in situ space exploration missions. Laser-based Mass Spectrometry techniques (LIMS, LD-MS) are compared with Laser-Induced Breakdown Spectroscopy (LIBS), Alpha Particle X-ray Spectrometry (APXS), and pyrolysis Gas Chromatography – Mass Spectrometry (pyr GC – MS). Typical sizes of biological units are indicated, as well as quantifities and concentrations required by space exploration missions or for element analysis. Note that the top and bottom sensitivity axis are related by the total sample mass. In both sensitivity and spatial resolution, laser-based instrumentation performs better than current space instruments.}
\label{fig:sensitivity} 
\end{figure}


\section{The Mars scenario}

While evidence points in the direction of a watery past for Mars, nowadays it is a mostly barren place, which has lost most of its liquid water. Therefore, the expectation is to find extinct life on Mars, most probably in the subsurface that protects the fragile signatures from the harsh ionizing environment. Recently, the most promising biosignatures were reviewed in detail and allocated to six groups, ranging from biomolecules to microstructures, to solids where specific elements are isotope fractionated (e.g., sulfur)\citep{hays2017}.
The high detection sensitivity of laser-based mass spectrometers (see Fig. 1 for a comparison of experimental techniques) coupled to spatially resolved chemical analysis, allows for the identification and chemical characterization of small scale features at the micrometer level. These advantages have now been recognized by the space exploration community and resulted in the inclusion of LD-MS in the Mars Organic Molecule Analyzer (MOMA) instrument on the ESA/Roscosmos ExoMars rover \citep{li2017,goesmann2017}. 
Recent measurement campaigns demonstrate that LIMS systems not only have the measurement capabilities to analyze the elemental composition of minerals \citep{neubeck2016}, lunar regolith \citep{managadze2010}, and meteorites \citep{frey2020} but can also identify putative fossil structures of micrometer dimensions embedded in a host material by monitoring elements relevant to life, such as carbon \citep{tulej2015,wiesendanger2018}. Measurements conducted on Martian mudstone analog material allowed the identification of the elemental signatures of single microbes that were inoculated artificially at low number density \citep{stevens2019}. Moreover, LIMS systems, such as LAb-CosmOrbitrap \citep{briois2016}, have been able to successfully characterize the chemical composition of organics and tholins mixed with minerals \citep{selliez2020,arevalo2018} making use of metal-organic clusters for identification and the high mass resolving power of Orbitrap to disentangle isobaric interferences of possible biogenic and non-biogenic molecular fragments. In combination with a microscope camera system, features of micrometer dimensions can be localized and targeted with the laser beam with high accuracy, which allows for the chemical investigation of the feature with minimal contribution of the host material that may obscure the faint life signature. Chemical imaging further refines the analysis of these microfossils by scanning the entire region of interest. 
Sample preparation is limited for LIMS measurements on Mars. In a best-case scenario, the LIMS system can be placed on a robotic arm to make chemical measurements on the local environment. Alternatively, the LIMS system can be placed on the payload platform on a lander or rover and have samples introduced to it via a sample delivery system. To conduct a thorough analysis, it needs to be possible to manipulate samples in the x,y,z direction and potentially rotate them. Note that the two laser-based mass spectrometry systems that to date have been prepared for space flight, LAZMA \citep{managadze2010} for the Roscosmos Phobos-Grunt (English: Phobos-Soil) mission, and MOMA \citep{goesmann2017} for the ESA/Roscosmos ExoMars rover, both have a sample carousel that allows for successive analyses of an aliquot of the sample across the arc of the carousel motion.


\section{The Europa scenario}

While Europa and other ocean worlds have inhospitable ice shells, they do contain subsurface oceans of liquid water and in some cases conditions that are considered right for life to emerge and persist \citep{lunine2017}. Through cracks in the ice crust or with plumes, signatures of this life can end up on its surface, form deposits, and become accessible for in situ analysis. However, the transfer of molecular biosignatures to the surface is likely inefficient and at the surface, these molecules can be processed by radiation. Instruments that search for these signatures thus need to be able to detect trace amounts of biomolecules and identify associations between detected organic compounds and geological phases that may shield them from harsh environmental conditions. LIMS and LD-MS instruments commonly reach picomol mm$^{-2}$ detection sensitivity \citep{goesmann2017,getty2012,arevalo2018,uckert2018,selliez2019,riedo2013a,riedo2013b,riedo2013c} and in some cases even achieve detection limits down to several femtomol mm$^{-2}$, such as with the LD-MS system called ORIGIN \citep{ligterink2020}. Space LIMS and LD-MS systems have demonstrated that they can detect several classes of biomolecules, such as amino acids and nucleobases, even in complex mixtures, or when mixed with minerals and salts \citep{briois2016,getty2012,arevalo2018,uckert2018,selliez2019,ligterink2020,moreno2016}. No signs for the chemical alteration of biomolecules by laser-induced reactions between minerals and organic molecules are seen during LIMS or LD-MS analysis of molecules mixed with minerals or salts. 
LIMS and LD-MS systems can perform measurements directly on ice samples, but this may not be the most efficient way to detect biomolecules. It is better to make use of the physical characteristics of Europa and its ice constituents, by simply warming up an ice sample until the water ice sublimates and a high concentration residue of biomolecules, salts, and minerals remain, which can be considered as a simple enrichment process. As shown before, these residues do not have to be filtered since the LIMS/LD-MS measurement does not chemically alter its components. Sample preparation is therefore limited and simple and only consists of warming up an ice sample after its introduction into an instrument, which is for free by using the thermal dissipation of the instrument payload \citep{ligterink2020}. 


\section{Other LIMS/LD-MS applications on space exploration missions}

Due to a large number of technical benefits and its measurement capabilities, LIMS and LD-MS instruments are an excellent addition to the scientific payloads of planetary exploration missions when searching for signatures of life. However, the unique properties of LIMS and LD-MS systems also make other uses possible. Since these systems are compact and lightweight they are suited for miniature landers or handheld applications as well. The Hayabusa-2 mission to asteroid Ryugu deployed multiple small landers, such as MASCOT, to investigate the asteroid in situ \citep{ho2017}. MASCOT weighs around 10 kg and has a volume of 15,500 cm$^{3}$. This type of lander can therefore easily fit a space LIMS/LD-MS instrument and be deployed to perform in situ chemical analysis of asteroids or comets. Due to their compact and lightweight nature, future space exploration missions could carry multiple miniature landers with LIMS/LD-MS systems and investigate multiple science targets in a single mission. Similarly, a handheld LIMS system can be used during future Moon landings to perform easy in situ analysis of Moon regolith. On sample return missions, LIMS and LD-MS can be employed to perform preliminary chemical analysis of sample sites, so the scientifically most interesting samples are collected and brought to Earth.


\section{Concluding remarks}

Laser-based mass spectrometer systems for use on space exploration missions have been developed and tested for over two decades, demonstrating the maturity of this technology. Due to the versatility of LIMS and LD-MS systems, they can be applied in a wide variety of applications where in situ chemical analysis is desired and are in particular well suited to find chemical indicators of life. Because chemical processing of samples is not required, LIMS/LD-MS instruments are compact, require no consumables, and above all are not biased to specific chemical groups. Chemical imaging and depth profiling on micrometer scales makes it possible to detect the chemical biosignatures of microfossils. A number of concept planetary exploration missions, such as the Europa Lander, Lunar landers, future Mars missions, and various sample return missions, can benefit from including a Laser Desorption/Ablation Mass Spectrometer to meet their scientific goals. 


\setlength{\bibsep}{0.0pt}


\section{List of co-signatories}

\indent Frances Westall
-- CNRS-Centre de Biophysique Moléculaire, Orléans, France
\\
Edith Fayolle	
-- Jet Propulsion Laboratory, USA
\\
Charles Cockell	
-- University of Edinburgh, UK
\\
Laurent Thirkel	-- 
Laboratoire de Physique et Chimie de l’Environnement et de l’Espace (LPC2E), Orléans, France
\\
Alexander Makarov	-- 
Thermo Fisher Scientific, Germany
\\
Vladimir Azov	-- 	
Department of Chemistry, University of the Free State, Republic of South Africa
\\
Xiang Li	-- 	
NASA Goddard Space First Center \& University of Maryland, USA
\\
Robert Lindner	-- 	
European Space Agency ESA-ESTEC, The Netherlands
\\
Cyril Szopa	-- 
University of Paris-Saclay, France
\\
Jorge Vago	-- 	
European Space Agency ESA-ESTEC, The Netherlands
\\
Adrian Southard	-- 
NASA Goddard Space First Center, USA

\clearpage

\end{document}